\documentclass[11pt]{article}
\usepackage{amsmath,amsthm,latexsym,amssymb,amsfonts,amsthm}
\usepackage{graphicx, color, calc}
\usepackage[list=true, font=large, labelfont=bf,
labelformat=brace, position=top]{subcaption}

\makeatletter \@addtoreset{equation}{section} \makeatother

\newtheorem{Definition}{Definition}[section]
\newtheorem{Lemma}{Lemma}[section]
\newtheorem{Corollary}{Corollary}[section]
\newtheorem{Proposition}{Proposition}[section]

\title{Volume of 4-polytopes from bivectors}
\author{Benjamin Bahr$^1$\\
\small $^1$ II Institute for Theoretical Physics\\
\small University of Hamburg\\
\small  Luruper Chaussee 149\\
\small 22761 Hamburg, Germany
 }
\date{}

\begin{document}

\maketitle

\begin{abstract}
In this article we prove a formula for the volume of 4-dimensional polytopes, in terms of their face bivectors, and the crossings within their boundary graph. This proves that the volume is an invariant of bivector-coloured graphs in $S^3$. 
\end{abstract}

\section{Introduction}

Loop Quantum Gravity and Spin Foam Models aim at a quantization of General Relativity, using discrete structures. \cite{thomasbook, cbook, Perez:2012wv, francescabook}\footnote{They share this feature with other, related approaches to quantum gravity  \cite{Ambjorn:2013apa, Bombelli:1987aa, Oriti:2007qd}}  In particular Spin Foam Models rest on the equivalence of GR with constrained BF theory \cite{Plebanski:1977zz, Horowitz:1989ng, Reisenberger:1996pu}. While the quantization of $BF$-theory is straightforward \cite{Baez:1999sr}, the connection to GR is implied by enforcing a version of the so-called simplicity constraints on the quantum theory. 

The discrete variables are parallel transport holonomies $g_\ell\in \text{Spin}(4)$ associated to $1d$ curves $\ell$, and bivectors $B_f\in \mathbb{R}^4\wedge \mathbb{R}^4$ associated to $2d$ surfaces $f$. These latter ones act as derivative operators on the boundary Hilbert space, and the asymptotic limit of the quantum amplitudes can describe transitions in Regge geometry of $4d$ polyhedra $P$ with given classical bivectors $B_f$ associated to their $2d$ subfaces \cite{Barrett:2009gg, Barrett:2009mw, Conrady:2008mk}.

Also the (discrete, classical version of the) simplicity constraints can be formulated in terms of the $B_f$, and they can be regarded as conditions  to allow a reconstruction of $P$ for given bivectors $B_f$. In that case the dynamics of the theory is intricately connected to the $4d$ geometry of space-time.

For the special case of a $4$-simplex, this reconstruction is known, and the corresponding simplicity constraints have been the foundation for modern spin foam models  \cite{Barrett:1997gw, Engle:2007wy, Freidel:2007py, CubulationSpinFoamThiemann2008, Baratin:2011hp}. While the formula can be extended to arbitrary polyhedra \cite{Kaminski:2009fm}, it still rests on the reconstruction of the 4-simplex, rather than that of the respective polyhedron (which in general is unknown). In particular, it could be shown that in the case of certain transitions, the corresponding model is underconstrained, and contains additional, non-geometric degrees of freedom. These can be directly traced to an insufficient implementation of the so-called quadratic volume simplicity constraint \cite{Bahr:2015gxa, Bahr:2017ajs, Belov:2017who}. \footnote{Although non-geometrc geometries manifest themselves in face-non-matching, they go beyond the so-called ``twisted geometries'', which are already present in the 4-simplex case \cite{Dittrich:2008va, Freidel:2010aq, Freidel:2013bfa}.}

An implementation of this constraint for the special case has been presented in \cite{Bahr:2017ajs}, and shown to reduce the system to the correct numbers of degrees of freedom. This constraint rests on a formula for the 4-volume of $P$ in terms of its bivectors $B_f$. In this article, we will give a proof of this formula for the case of general $P$. It rests, beyond the bivectors $B_f$, on the knotting class of the boundary graph $\Gamma$ of $P$. In particular, for $\Gamma$ projected onto the plane with crossings $C$, the volume $V_P$ of a (convex) polytope $P$ can be expressed as
\begin{eqnarray}\label{Eq:CentralFormula}
V_P\;=\;\frac{1}{6}\sum_C\sigma(C) *(B_1\wedge B_2),
\end{eqnarray}

\noindent where $\sigma(C)$ is the sign of the crossing, $B_1$, $B_2$ are the bivectors associated to the graph links involved in the crossing (which correspond to $2d$ faces in a prospective $P$), and $*$ is the Hodge dual. This formula was also used in \cite{Bahr:2018ewi, Bahr:2018gwf} to add a cosmological-constant-like deformation to the Spin Foam Model\footnote{The 4-simplex-version of this operator was defined in \cite{Han:2011aa} and used in \cite{Haggard:2014xoa} to define a relation to Chern-Simons theory of a deformed amplitude. A different but related way to include a cosmological constant in the amplitude is by replacing classical with quantum groups \cite{Fairbairn:2010cp, Han:2010pz}, which can also be done on the boundary Hilbert space level \cite{Dittrich:2016typ}.}. In the following, we will prove the formula for general convex polytopes in $4d$.

The strategy of the proof is as follows: First we define an invariant $I(\Gamma)$, as the rhs of (\ref{Eq:CentralFormula}), for arbitrary graphs $\Gamma$, hence for arbitrary convex polytopes. Then we show that, when a polytope is cut into two smaller convex polytopes by a hyperplane,  the sum of the invariants of the new polytopes add up to the invariant of the old one. Then, we show that in case of a 4-simplex, the value of the invariant coincides precisely with its 4-volume, i.e.~(\ref{Eq:CentralFormula}) is true for 4-simplices. Lastly, we show that every polyhedron can be successively cut by hyperplanes, until only 4-simplices remain. This proves (\ref{Eq:CentralFormula}) for arbitrary convex polytopes.

\section{Bivector-coloured graphs}

\begin{Definition}
Let $\Gamma$ be a directed graph embedded in $S^3$. Denote the set of nodes and (oriented) links of $\Gamma$ by $N(\Gamma)$ and $L(\Gamma)$. We call a \emph{bivector-colouring} of $\Gamma$ a map 
\begin{eqnarray}
B\,:\,L(\Gamma)\,\to\,\mathbb{R}^4\wedge \mathbb{R}^4. 
\end{eqnarray}

\noindent If the link $\ell$ meets the node $n$, we write $\ell\supset n$. For $\ell\supset n$, we write $[n,\ell]=1$, if the link $\ell$ is incoming but not outgoing, $-1$ if it is outgoing but not incoming, and $0$ if it is both or neither. We call a bivector-colouring of $\Gamma$ \emph{simple}, if the $B_\ell$, for $\ell\in L(\Gamma)$, satisfy the following conditions:
\begin{itemize}
\item For $\ell,\ell'$ both meeting at $n$, we have 
\begin{eqnarray}\label{Eq.SimplicityConstraints}
B_\ell\wedge B_{\ell'}\;=\;0.
\end{eqnarray}
\noindent Note that this includes the case $\ell=\ell'$. 
\item For every node in $\Gamma$ we have that
\begin{eqnarray}\label{Eq:ClosureCondition}
\sum_{\ell\supset n}[n,\ell]B_\ell\;=\;0.
\end{eqnarray}
\end{itemize}
\end{Definition}

In the spin foam literature, conditions (\ref{Eq.SimplicityConstraints}) are called ``diagonal simplicity''for $\ell=\ell'$ and ``cross-simplicity'' for $\ell\neq\ell'$, while (\ref{Eq:ClosureCondition}) is called the ``closure condition''. In the following, we denote, for simplicity, a simple bivector-coloured graph as $\Gamma$, rather than $(\Gamma,\{B_\ell\}_{\ell\in L(\Gamma)})$. 

\begin{figure}[hbt!]
\begin{minipage}[t]{0.45\textwidth}
\begin{center}
\includegraphics[scale = 0.45]{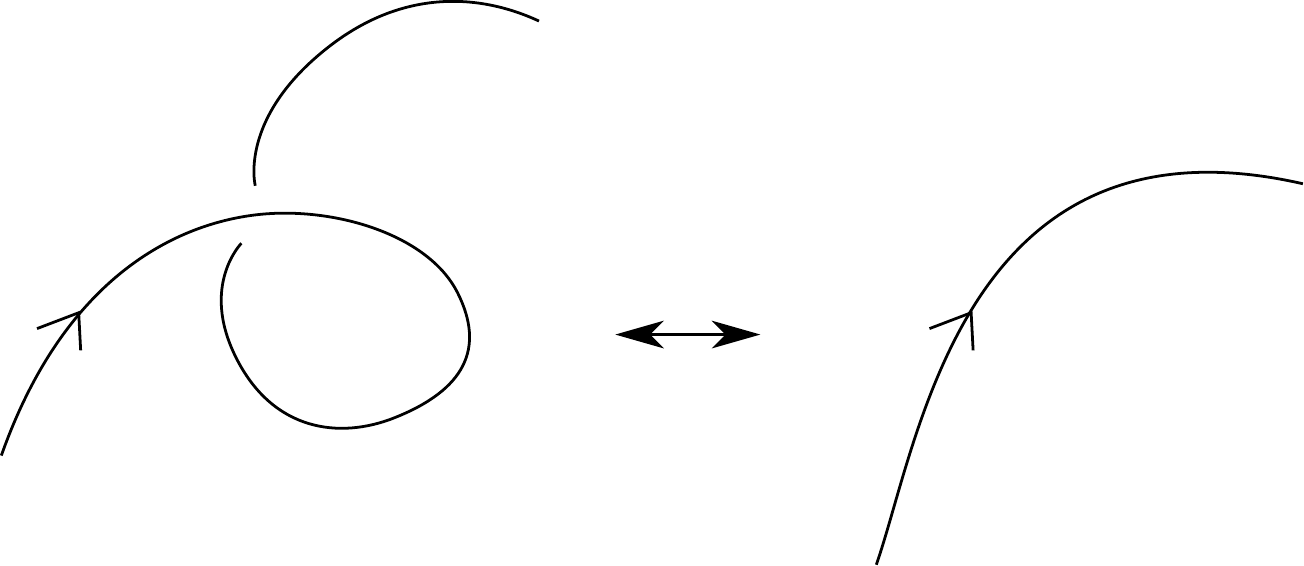}
\caption{Removing a writhing\\ (self-crossing of a link).}\label{Fig:WrithingRemoved}
\end{center}
\end{minipage}
\begin{minipage}[t]{0.45\textwidth}
\begin{center}
\includegraphics[scale = 0.45]{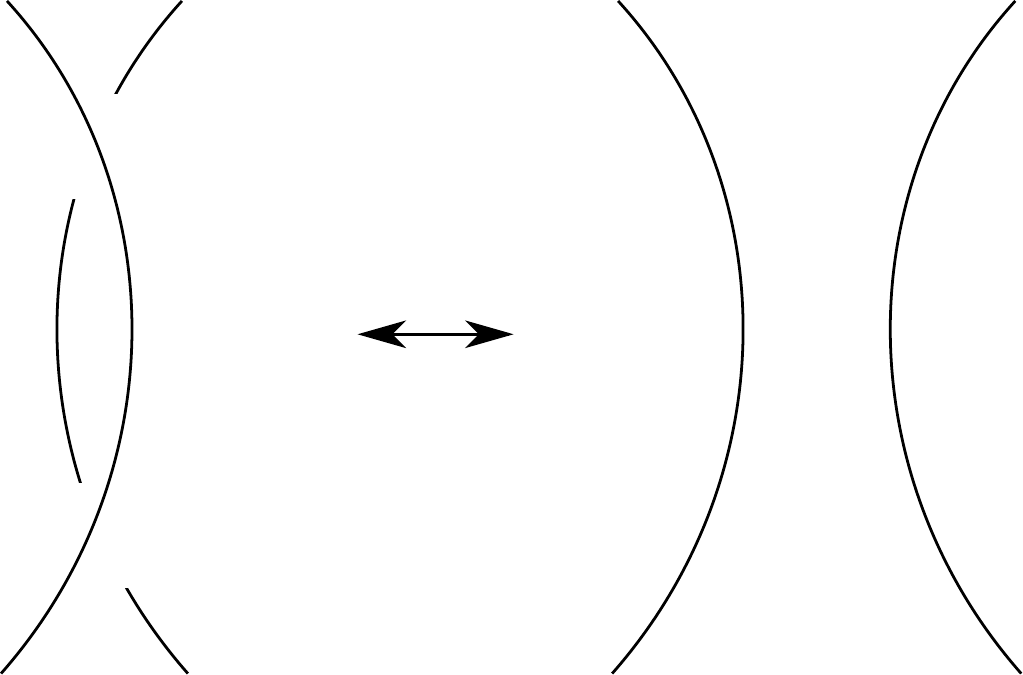}
\caption{Sliding two links across one another.}\label{Fig:SlidingOff}
\end{center}
\end{minipage}
\end{figure}

\begin{figure}[hbt!]
\begin{center}
\includegraphics[scale = 0.5]{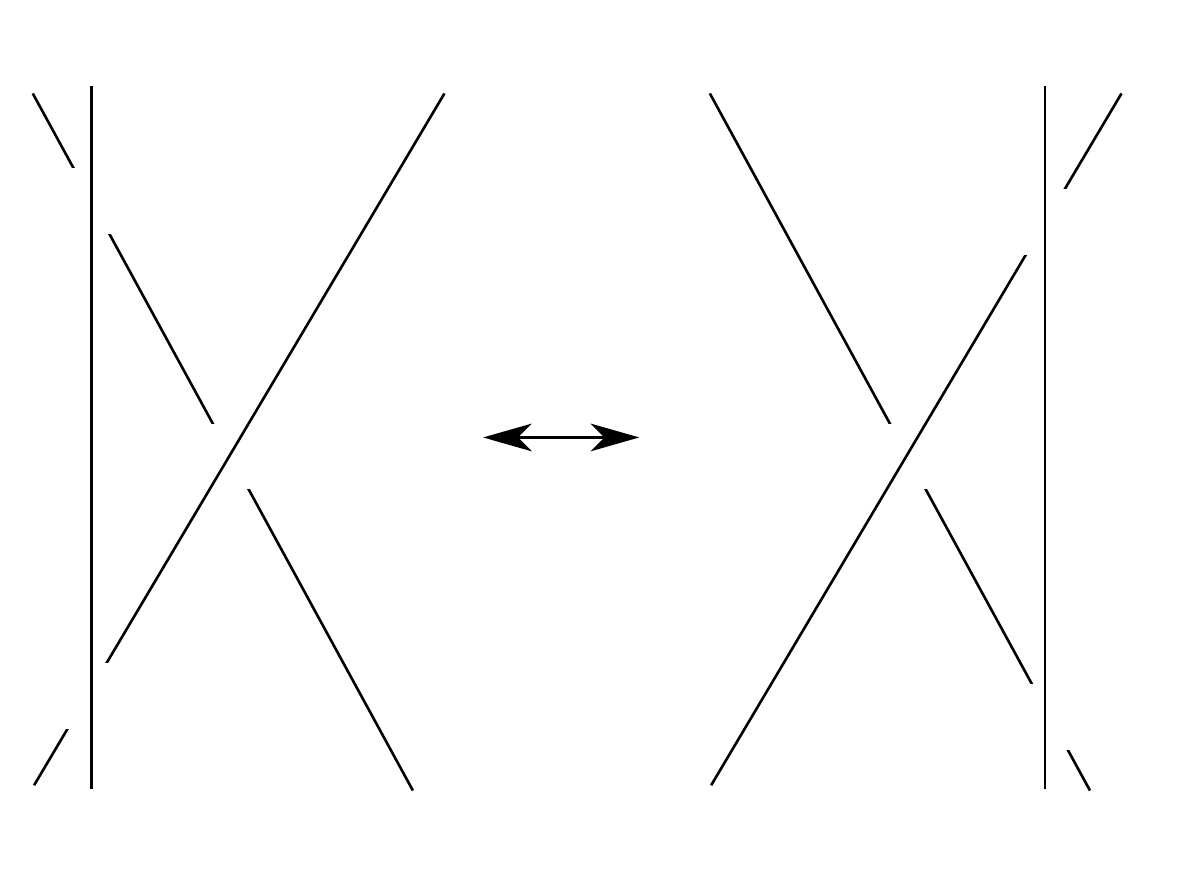}
\caption{Sliding links under / over crossings.}\label{Fig:KnottingMove}
\end{center}
\end{figure}

\begin{figure}[hbt!]
\begin{minipage}[t]{0.45\textwidth}
\begin{center}
\includegraphics[scale = 0.45]{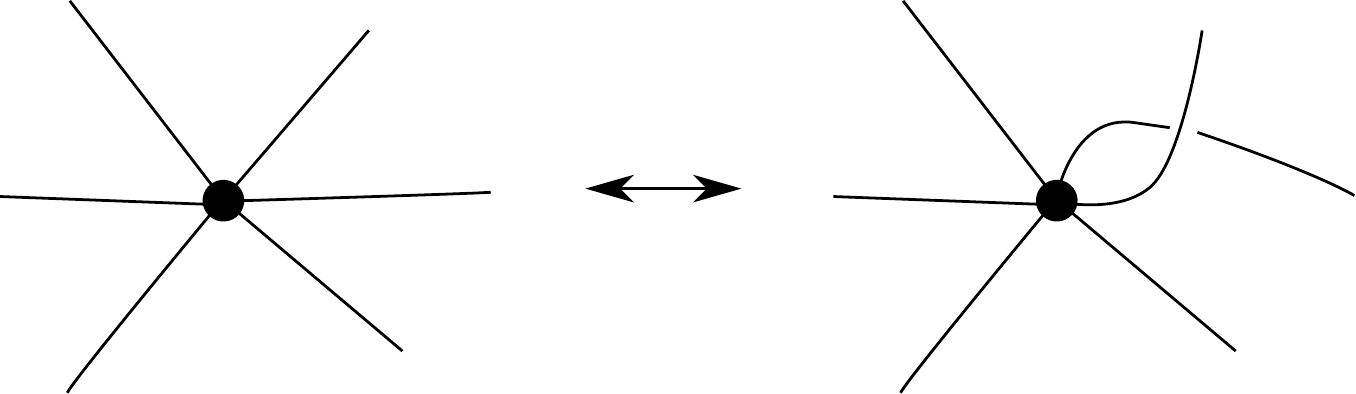}
\caption{Twisting links at nodes.\\This changes the cyclic ordering of\\ links.}\label{Fig:NodeTwisting}
\end{center}
\end{minipage}
\begin{minipage}[t]{0.45\textwidth}
\begin{center}
\includegraphics[scale = 0.45]{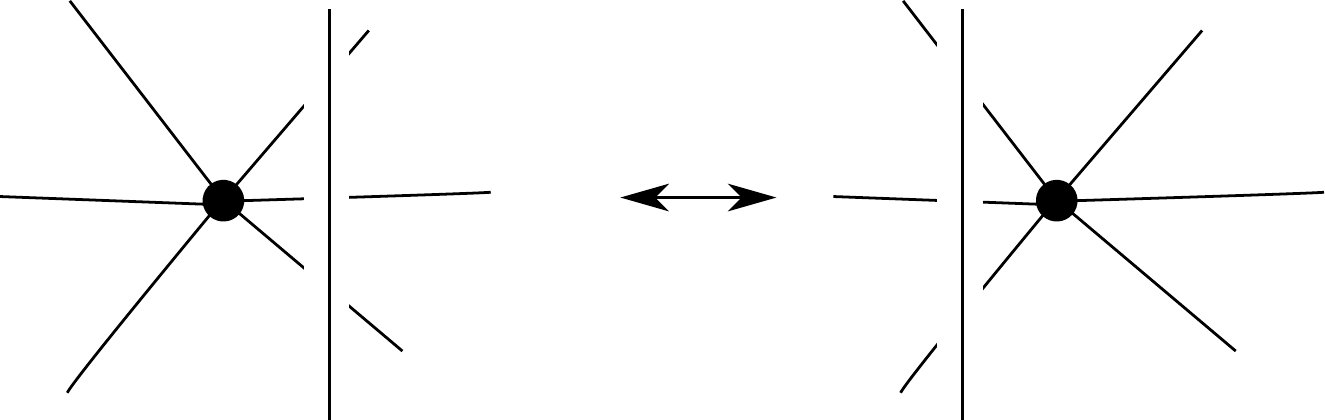}
\caption{Sliding links under / over nodes.}\label{Fig:CrossNodes}
\end{center}
\end{minipage}
\end{figure}

We call a projection of $\Gamma$ onto the plane a representation as graph $\tilde{\Gamma}$ on $\mathbb{R}^2$ with crossings. A projection can be achieved by choosing a point $p\in S^3$ which does not lie on $\Gamma$, then making a stereographic projection $\phi:S^3\backslash\{p\}\to\mathbb{R}^3$ w.r.t.~that point, to obtain a graph $\phi(\Gamma)$ in a compact subset of $\mathbb{R}^3$. A projection is then achieved into some direction such that no nodes lie on top of links, and links only cross one another finitely many times. Any two projections of $\Gamma$ can be obtained by the following moves on $\tilde{\Gamma}$ \cite{Reidemeister_Knots,AlexanderBriggs, GrossTuckerGraphBook}:
\begin{enumerate}
\item Remove writhing: figure \ref{Fig:WrithingRemoved}.
\item Link sliding over/under links: figure \ref{Fig:SlidingOff}.
\item Link sliding over/under crossings: figure \ref{Fig:KnottingMove}.
\item Links twisting at nodes: figure \ref{Fig:NodeTwisting}.
\item Link sliding over/under nodes: figure \ref{Fig:CrossNodes}.
\end{enumerate}

\begin{figure}[hbt]
\begin{center}
\includegraphics[scale = 1.0]{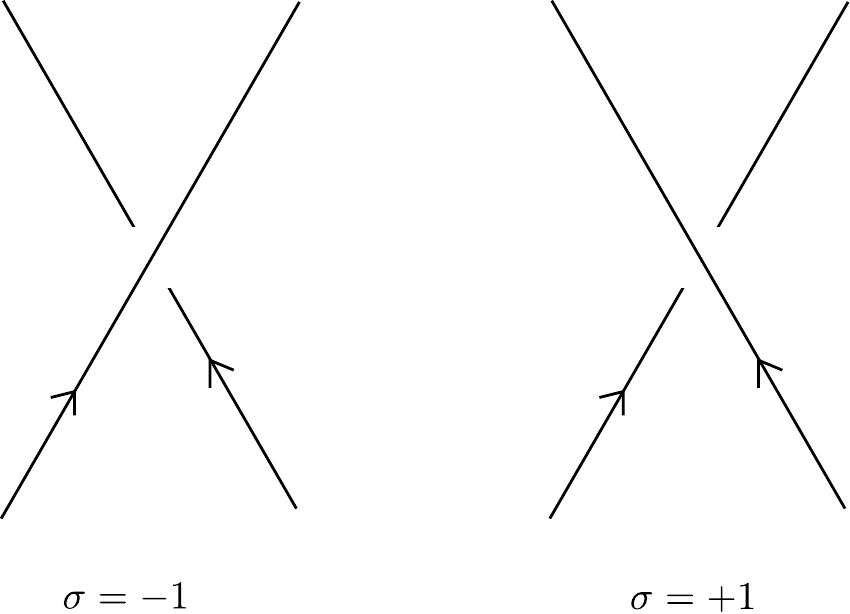}
\caption{The two possible knotting cases, which get assigned $\sigma = -1$ and $\sigma = +1$, respectively. Note how in both cases there is a well-defined \emph{upper} and \emph{lower} link.}\label{Fig:TwoKnottingCases}
\end{center}
\end{figure}

\begin{Definition}
Let $\Gamma$ be a simple bivector-coloured graph. For a projection $\tilde{\Gamma}$, there are two types of crossings, depending on the relation of link orientations and over-/undercrossings. These are depicted in figure \ref{Fig:TwoKnottingCases}. For a crossing $C$, define $\sigma(C)=\pm1$ depending on which of the two cases are present. Then we define the number
\begin{eqnarray}\label{Eq:DefinitionOfInvariant}
I(\Gamma)\;:=\;\frac{1}{6}\sum_{C}\sigma(C)\;*\left(B_{\ell_1}\wedge B_{\ell_2}\right).
\end{eqnarray}

\noindent Here $*:\wedge^4\mathbb{R}^4\to\mathbb{R}$ is the Hodge dual, and $\ell_1$ and $\ell_2$ are the two edges participating at the crossing $C$. Since for bivectors $B$, $B'$, one has that $B\wedge B'=B'\wedge B$, it is not important which of the $\ell_1$, $\ell_2$ is the upper or lower link. The sum ranges over all crossings of the projection $\tilde{\Gamma}$. 
\end{Definition}

\noindent The notation $I(\Gamma)$ instead of $I(\tilde{\Gamma})$ can be explained with the following proposition.
\begin{Proposition}\label{Prop:InvarianceUnderEquivalenceMoves}
The number $I(\Gamma)$ is invariant under the moves described above.
\end{Proposition}
\noindent\textbf{Proof:} a) follows directly from $B_\ell\wedge B_\ell=0$, while b) follows from the fact that in any pair of over- or undercrossings $C_1$ and $C_2$, the two generated crossings are of opposite type, i.e.~$\sigma(C_1)=-\sigma(C_2)$, while the participating bivectors are the same in both crossings. c) follows from (\ref{Eq:ClosureCondition}), while d) follows from (\ref{Eq.SimplicityConstraints}) for $\ell\neq\ell'$. \qed

Indeed, the number $I(\Gamma)$ only depends on the graph embedded in $S^3$ (up to ambient isotopy) and its bivector colouring, not on any choice of projection onto the plane with crossings. 

%Another crucial observation is that the number $I(\Gamma)$ is invariant under a change of the orientation of one link $\ell$, as long as $B_\ell\to-B_\ell$ at the same time. This will be important in the following, when we might choose to switch orientations of links, with complementary sign changes of the associates bivector, when it is convenient for our proofs. 

\begin{figure}[hbt]
\begin{center}
\includegraphics[scale = 0.5]{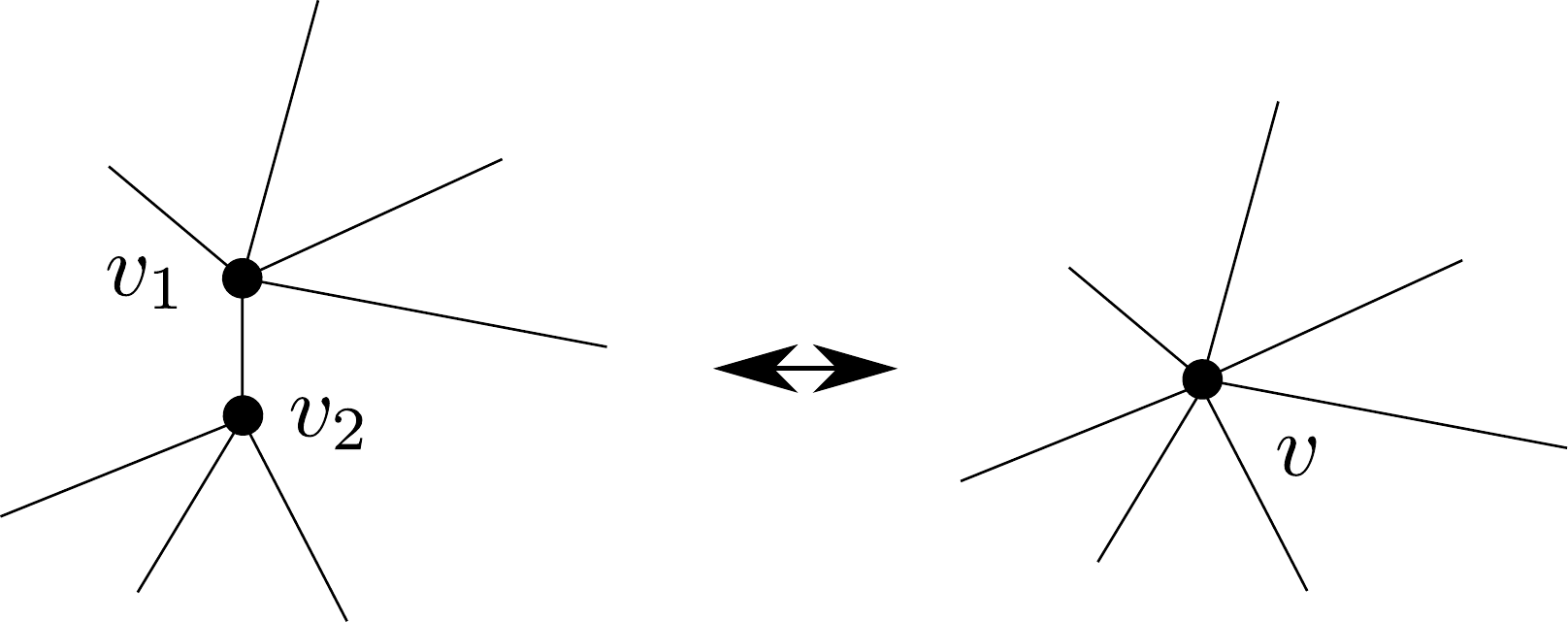}
\caption{The merging of two nodes $v_1$, $v_2$ to $v$ is only allowed, if all links meeting at either $v_i$ satisfy (\ref{Eq.SimplicityConstraints}).}\label{Fig:MergeNodes}
\end{center}
\end{figure}

\begin{figure}[hbt]
\begin{center}
\includegraphics[scale = 0.45]{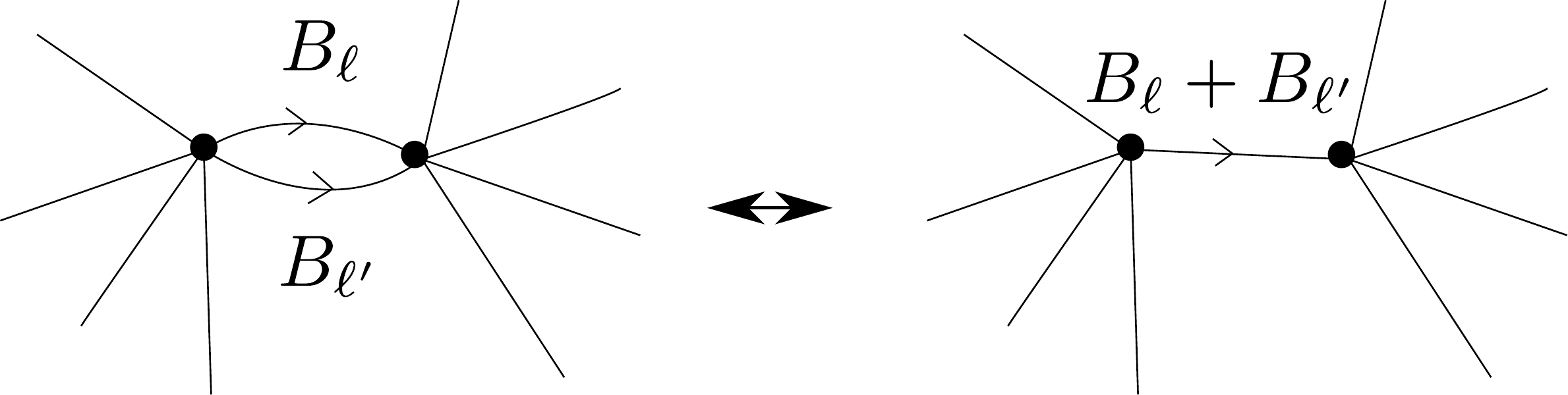}
\caption{The merging of two links is always allowed.}\label{Fig:MergeLinks}
\end{center}
\end{figure}

\begin{Definition}\label{Def:GraphEquivalence}
Let $\Gamma$, $\Gamma'$ be two bivector-coloured graphs embedded in $S^3$. Denote $\Gamma\sim\Gamma'$, if one arises from the other by a finite sequence of the following moves (or their inverses):
\begin{enumerate}
\item A continuous ambient isotopy, i.e.~a homeomorphism of $S^3$ to itself,
\item A change of orientation of a link $\ell$, accompanied by a change of $B_\ell\to -B_\ell$,
\item In case that there are two nodes $v_{1}$ and $v_2$, and a link $\ell\supset v_{1,2}$ between them, such that all pairs of bivectors $B_\ell$, $B_{\ell'}$ for links touching either of the two nodes satisfy $B_\ell\wedge B_{\ell'}=0$: A merging of the two nodes, i.e.~remove both $v _{1,2}$ and link $\ell$, and replace it with another node $v$ in their vicinity, with all other links connected to $v$ (see figure \ref{Fig:MergeNodes}).
\item In case that there are two nodes $v_1$, $v_2$ and two similarly-oriented links $\ell_{1,2}\supset v_{1,2}$, such  that there is an ambient isotopy which moves $\ell_1$ to $\ell_2$, but leaves all $\ell\neq\ell_{1,2}$ untouched: A merging of those two links, i.e.~remove $\ell_2$, and replace $B_{\ell_1}$ by $B_{\ell_1}+B_{\ell_2}$ (see figure \ref{Fig:MergeLinks}). 
\end{enumerate}
\end{Definition}

\begin{Proposition}
The moves prescribed in definition \ref{Def:GraphEquivalence} transform simple bivector-coloured graphs to simple bivector-coloured graphs. 
\end{Proposition}
\noindent\textbf{Proof:} It is straightforward to show that neither 1.), 2.) nor 4.) change (\ref{Eq.SimplicityConstraints}) or (\ref{Eq:ClosureCondition}). It is also clear that after a merging of nodes the new node satisfies (\ref{Eq:ClosureCondition}), since the removed link was ingoing into one, and outgoing of the other original node. By the condition, after the merging of nodes, the condition (\ref{Eq.SimplicityConstraints}) holds for all new links at the new node, so 3.) also generates a simple bivector-coloured graph. 

\begin{Proposition}\label{Prop:EquivalentGraphsHaveEqualInvariants}
Let $\Gamma\sim\Gamma'$ be two bivector-coloured graphs, then $I(\Gamma)=I(\Gamma')$. 
\end{Proposition}
\noindent\textbf{Proof:}  By proposition \ref{Prop:InvarianceUnderEquivalenceMoves}, 1.) is clear. 3.) is also clear, since in any projection, we can make moves (fig.~\ref{Fig:WrithingRemoved} -- fig.~\ref{Fig:CrossNodes}), until $\ell$ does not partake in any crossing. If the conditions for 4.) are satisfied, one can find a projection such that neither $\ell_1$ nor $\ell_2$ partake in a crossing, so 4.) is true as well. To show invariance under 2.), just note that for any crossing $C$ between two different links, changing orientation of one of the links $\ell$ changes the sign of $\sigma(C)$. The accompanying sign change of $B_\ell$ compensates for it, so $I(\Gamma)$ is unchanged. \qed

\begin{figure}[hbt]
\begin{center}
\includegraphics[scale = 0.35]{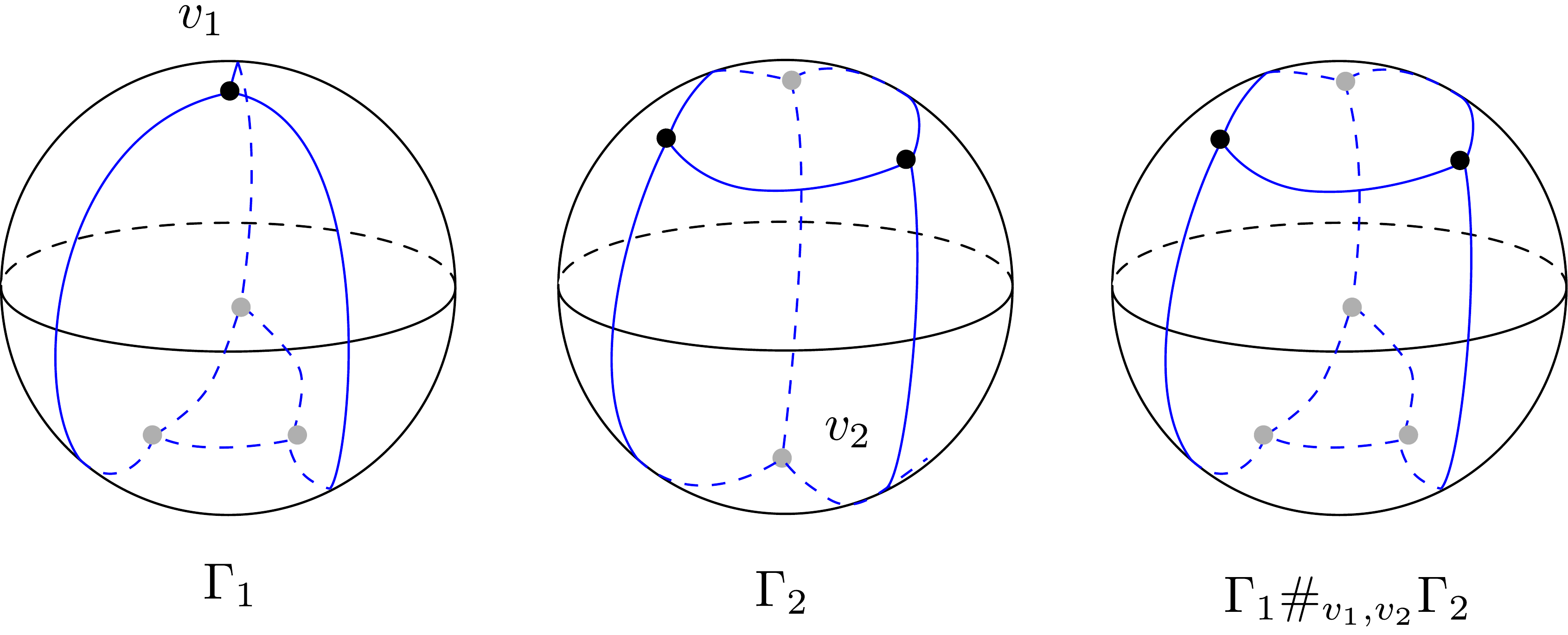}
\caption{Two graphs $\Gamma_1$ and $\Gamma_2$ can be glued along two nodes, if the bivectors around those nodes agree and the link orientations are opposite. Figure shows the $3d$ analogue of the $4d$ procedure.}\label{Fig:MergeGraphs}
\end{center}
\end{figure}

\begin{Definition}
Consider two bivector-coloured graphs $\Gamma_i$, $i=1,2$. Assume that there are two nodes $v_i\in N(\Gamma_i)$, $i=1,2$, which have the following property: Around both, respectively exist open, simply-connected neighbourhoods $U_i\subset S^3$ containing $v_i$ as the only node, and homeomorphisms $\phi_i$ in $S^3$ which map $U_i$ to the northern (for $i=1$) and southern (for $i=2$) hemisphere of $S^3$. Furthermore, let $\phi_i$ map $\partial U_i$ to the equator $S^2\subset S^3$, and  $v_i$ onto the north pole ($i=1$) and south pole ($i=2$), and the respective link segments onto geodesic lines from the respective pole to the equator (see figure \ref{Fig:MergeGraphs}).

Finally assume there is a one-to-one correspondence between links $\ell\supset v_1$ and links $\ell'\supset v_2$ such that 
\begin{itemize}
\item To every link $\ell\supset v_1$ corresponds an $\ell'\supset v_2$ with $[v_1,\ell]=-[v_2,\ell']$.
\item For every two such corresponding link pairs one has $B_\ell=B_{\ell'}$. 
\item For every such pair $\phi_1(\ell\cap \partial U_1)=\phi_2(\ell'\cap\partial U_2)$ is a point lying in $S^2$. 
\end{itemize} 
\noindent If these conditions are satisfied, we say that $\Gamma_1$ and $\gamma_2$ can be glued together (at $v_1$ and $v_2$). In that case, we can define a new bivector-coloured graph, denoted by $\Gamma_1\#_{v_1,v_2}\Gamma_2$, which is such that 
\begin{eqnarray*}
\left(\Gamma_1\#_{v_1,v_2}\Gamma_2\right)\cap N\;&=&\;\phi_1(\Gamma_1)\cap N,\\[5pt]
\left(\Gamma_1\#_{v_1,v_2}\Gamma_2\right)\cap S\;&=&\;\phi_2(\Gamma_2)\cap S,
\end{eqnarray*}
\noindent where $N$, $S$, are the northern and southern hemisphere of $S^3$, respectively.
\end{Definition}

\noindent Figure \ref{Fig:MergeGraphs} shows a 3-dimensional analogue the construction. In essence, two bivector-coloured graphs which have ``opposite nodes'', i.e.~which have two nodes with the same incident links and bivectors on those links, but negative relative orientation, can be ``glued together'' along those two nodes. If both $\Gamma_i$, $i=1,2$ are simple, then so is $\Gamma_1\#_{v_1,v_2}\Gamma_2$, of course. 

The concept is analogous to the well-known procedure of surgery from differential geometry. However, the condition on the two nodes is quite severe, and it may well be that two graphs cannot be glued together like this at all. 

\begin{Proposition}\label{Prop:InvariantBehaviourUnderGlueingOfPolyhedra}
For any two simple bivector-coloured graphs $\Gamma_1$ and $\Gamma_2$ which can be glued along $v_1$ and $v_2$, one has
\begin{eqnarray}
I\left(\Gamma_1\#_{v_1,v_2}\Gamma_2\right)\;=\;I(\Gamma_1)\,+\,I(\Gamma_2).
\end{eqnarray}
\end{Proposition}
\noindent\textbf{Proof:} One can find (by using ambient isotopies, or projecting from points close to the north pole) projections such that $\phi_1(\Gamma_1\cap U_1)$ is being projected to $\mathbb{R}^2\backslash D$ (where $D$ is the solid circle with radius $1$), with no crossings inside $D$. Also, $\phi_2(\Gamma_2\cap U_2)$ can be mapped to $D$, where the only crossings outside of $D$ are among bivectors $B_1$, $B_2$ with $B_1\wedge B_2=0$, due to simplicity. The claim then follows due to the additivity of (\ref{Eq:DefinitionOfInvariant}) as sum over crossings.

\section{Convex polytopes in $d=4$.}

Convex polytopes $P$ in $\mathbb{R}^4$ can be represented via a set of half-spaces $H_i\;=\{x\in\mathbb{R}^4:\langle x,u_i\rangle \leq 1\}$, with $u_i\in\mathbb{R}^4$, via
\begin{eqnarray}
P\;=\;\bigcap_{i=1}^nH_i.
\end{eqnarray}
\noindent We assume that the presentation is irreducible, i.e.~none of the $H_i$ can be removed without changing $P$. In this case, there is a one-to-one correspondence between half-spaces $H_i$ and $3$-dimensional faces 
\begin{eqnarray}
\tau_i\;=\;P\cap \partial H_i.
\end{eqnarray}
\noindent Each of the 3-faces $\tau_i$ itself is a convex 3-dimensional polytope lying in the 3-dimensional affine subspace $\partial H_i$. Two neighbouring 3-faces can touch at a common 2-face $f$, in which case
\begin{eqnarray}
f\;=\;\tau_i\cap\tau_j\;=\;P\cap \partial H_i\partial H_j
\end{eqnarray}  
\noindent is a convex 2-dimensional polygon. 
\begin{Definition}
The (boundary) graph $\Gamma_P$ of a 4-polytope $P$ is a graph embedded in $S^3$, defined the following way: Without loss of generality, let the origin of $\mathbb{R}^4$ lie inside of $P$. Consider the sphere $S^3$ with radius 1, and project every point on $\partial P=\cup_i\tau_i$ onto $S^3$, via 
\begin{eqnarray}
\pi:\partial P\to S^3,\qquad \pi(x)\;\to\;\frac{x}{|x|}.
\end{eqnarray}
\noindent Choose a point $v_i$ in  each $\pi(\tau_i)$. These are the nodes of $\Gamma_P$. For two neighbouring $\tau_i$, connect the corresponding $v_i$ with a geodesic arc in $S^3$, such that it passes through $\pi(f)$. These are the links $\ell$ of $\Gamma_P$.  
\end{Definition}

\begin{figure}[hbt]
\begin{center}
\includegraphics[scale = 0.45]{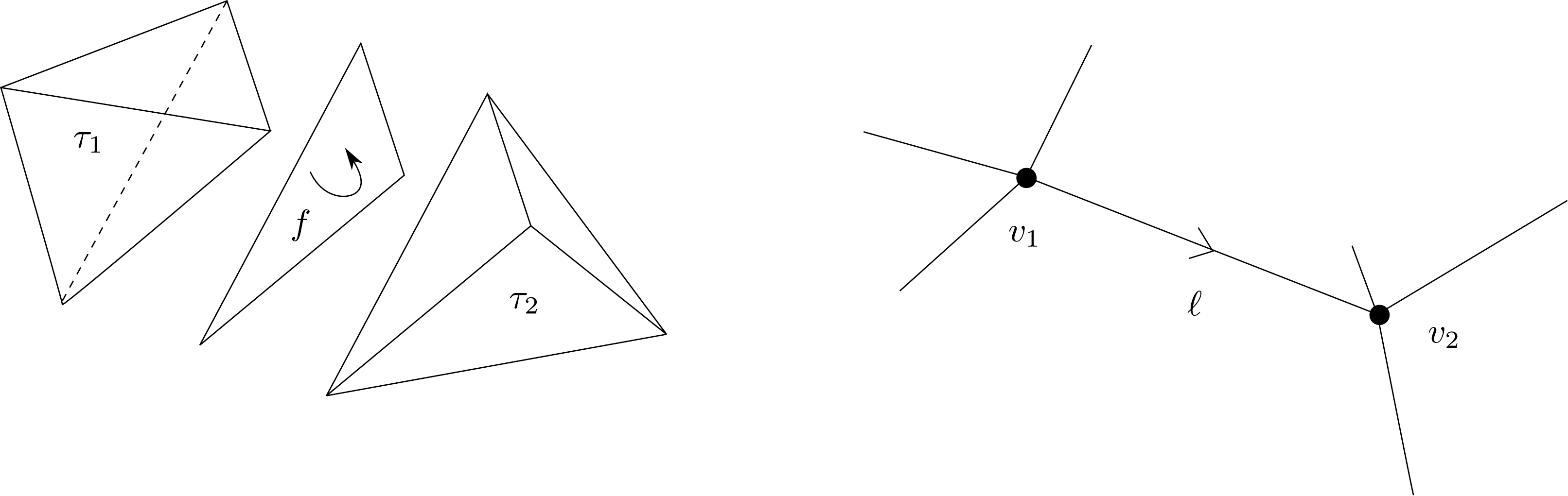}
\caption{An orientation of a face $f$ in $P$ determines the orientation of the link $\ell$ dual to it in $\Gamma_P$, and vice versa.}\label{Fig:Orientation}
\end{center}
\end{figure}

\noindent The thus constructed graph is, of course, not unique. But different choices are equivalent under homeomorphisms of $S^3$. Note that each link $\ell$ in $\Gamma_P$ is in one-to-one correspondence with 2-faces $f$ in $P$. Also note that $\Gamma_P$ does not come with an orientation for its links, but a choice of such is equivalent to an orientation of the corresponding $f$ in the following way: Each orientation of $f\subset P$ is given by a non-vanishing $2$-form $\omega_f$, which can be pulled back to a $2$-form defined on $\pi(f)\subset S^3$. Using the standard metric on $S^3$, we can convert this via musical isomorphism and Hodge duality to a non-vanishing normal vector field $X$ on $\pi(f)$. Then orient $\ell$ such that $X$ points from the source 3-polytope of $\ell$ to its target 3-polytope (see figure \ref{Fig:Orientation}).

\begin{Definition}
For a 2-face $f\subset P$, and a chosen orientation $\omega_f$ on it, define the (unique) bivector 
\begin{eqnarray}
B_f\;:=\;N\wedge M,
\end{eqnarray}
\noindent where $N,M\in\mathbb{R}^4$ are vectors that span $f$, such that $\omega_f(B_f)>0$ and $|B_f|=|N||M|\sin\theta$, where $\theta$ is the angle between $N$ and $M$. For a given choice of orientation of all faces, this defines an orientation of links of $\Gamma_P$, and, due to the one-to-one correspondence of faces $f$ and links $\ell$, a bivector-colouring of $\Gamma_P$. This is also called the bivector geometry of $P$. 
\end{Definition}

\begin{Proposition}
For a convex 4-polytope, the bivector geometry is a simple bivector-colouring of $\Gamma_P$. 
\end{Proposition}
\noindent\textbf{Proof:} The simplicity condition (\ref{Eq.SimplicityConstraints}) follows from the fact that, for two faces $f_{1,2}\subset\tau$, the bivectors satisfy $B_{f_1} \sim N_1\wedge M_1$, and $B_{f_2}\sim N_2\wedge M_2$. But all four vectors $N_{1,2},M_{1,2}$ lie in the $3d$ subspace parallel to the $3d$ polyhedron $\tau$, hence they are linearly dependent. Thus
$B_{f_1}\wedge B_{f_2}\sim N_1\wedge M_1\wedge N_2\wedge M_2 = 0$. 

To prove the closure condition (\ref{Eq:ClosureCondition}), we consider w.l.o.g.~the case that all links are outgoing of $n$. Denote the area of $f$ by $a_f$, and choose $N_f$ and $M_f$ as two orthogonal normalized vectors lying in the plane parallel to the face $f$, such that $B_f = a_f \,N_f\wedge M_f$ for all $f\subset\tau$. Denote by $T$ the normalized vector orthogonal to $\tau$, and for each $f$ let $O_f$ be the normalized vector in the $3d$ space parallel to $\tau$ orthogonal to $N_f$ and $M_f$, such that $(T, N_f, M_f, O_f)$ is positively oriented. Minkowski's theorem \cite{MinkowksiTheorem} states that 
\begin{eqnarray}
\sum_{f\subset\tau}a_f\,O_f\;=\;0.
\end{eqnarray}

\noindent Since $*(T\wedge O_f)=N_f\wedge M_f$, the claim follows.\qed

\begin{Proposition}\label{Prop:GraphBehaviourUnderGlueingOfPolyhedra}
Let $P$ be a 4-polytope, and $H\subset\mathbb{R}^4$ a half-space such that $P_1:=H\cap P$, $P_2:=(-H)\cap P$ are two 4-polytopes. Here $(-H):=\overline{\mathbb{R}^4\backslash H}$. If $v_1$ is the node of $\Gamma_{P_1}$ corresponding to the 3-face $\partial H\cap P_1$, and $v_2$ the node corresponding to the 3-face $\partial(-H)\cap P_2$, then
\begin{eqnarray}\label{Eq:GraphBehaviourUnderGlueingOfPolyhedra}
\Gamma_{P_1}\#_{v_1,v_2}\Gamma_{P_2}\;\sim\;\Gamma_P.
\end{eqnarray} 
\end{Proposition}

\noindent\textbf{Proof:} First we characterise the boundary graphs $\Gamma_{1,2}$. The polytope $P$ has a representation as intersection of half-spaces as
\begin{eqnarray}
P\;=\;H_1\cap H_2\cap\cdots\cap H_n.
\end{eqnarray}

\noindent Corresponding to these, we numerate the 3-faces of $P$ as $\tau_i=\partial H_1\cap P$. Intersecting this with $H$ and $(-H)$, respectively, one can w.l.o.g.~find an irreducible representation in terms of 
\begin{eqnarray*}
P_1\;&=&\;H\cap H_1\cap\cdots \cap H_k\cap H_{k+1}\cap\cdots \cap H_m,\\[5pt]
P_2\;&=&\;(-H)\cap H_1\cap \cdots \cap H_k\cap H_{m+1}\cap \cdots \cap H_n,
\end{eqnarray*}

\noindent with $k\leq m\leq m$. Note that it is possible that $k=m$, but not that $m=n$. The boundary 3-faces $\tau_1,\ldots\tau_k$ of $P$ are therefore the ones which get separated into two 3-faces by $\partial H$, and thus also appear in both $P_1$ and $P_2$. W.l.o.g.~we assume that the half-space $H$ is $H=\{x\in\mathbb{R}^4\,|\,x^1\geq 0\}$, i.e.~$(-H)=\{x\in\mathbb{R}^4\,|\,x^1\leq 0\}$. Furthermore, for $\tau_i$ and $\tau_j$ sharing a common face $f$ with with $k<i\leq m$ and $m<j\leq n$, that face $f$ has a corresponding link $\ell$ which connects a node in the northern and southern hemisphere, i.e. crosses the equator $E\simeq S^2\subset S^3$. We furthermore assume w.l.o.g.~that each node corresponding to a 3-face $\tau_i$, $1\leq i\leq k$, lies on $E$, as well as any link between two such neighbouring nodes.

The 3-face $\tau=\partial H\cap P_1$ in $P_1$ has as 2-faces those which are on the boundary of $\tau$, i.e.~those which are either intersections of $\tau_i\cap \partial H$, with $1\leq i\leq k$, or those which are faces between $\tau_i$ and $\tau_j$, with $k<i\leq m$, $m<j\leq n$.

\begin{figure}[hbt!]
\begin{center}
\includegraphics[scale = 0.45]{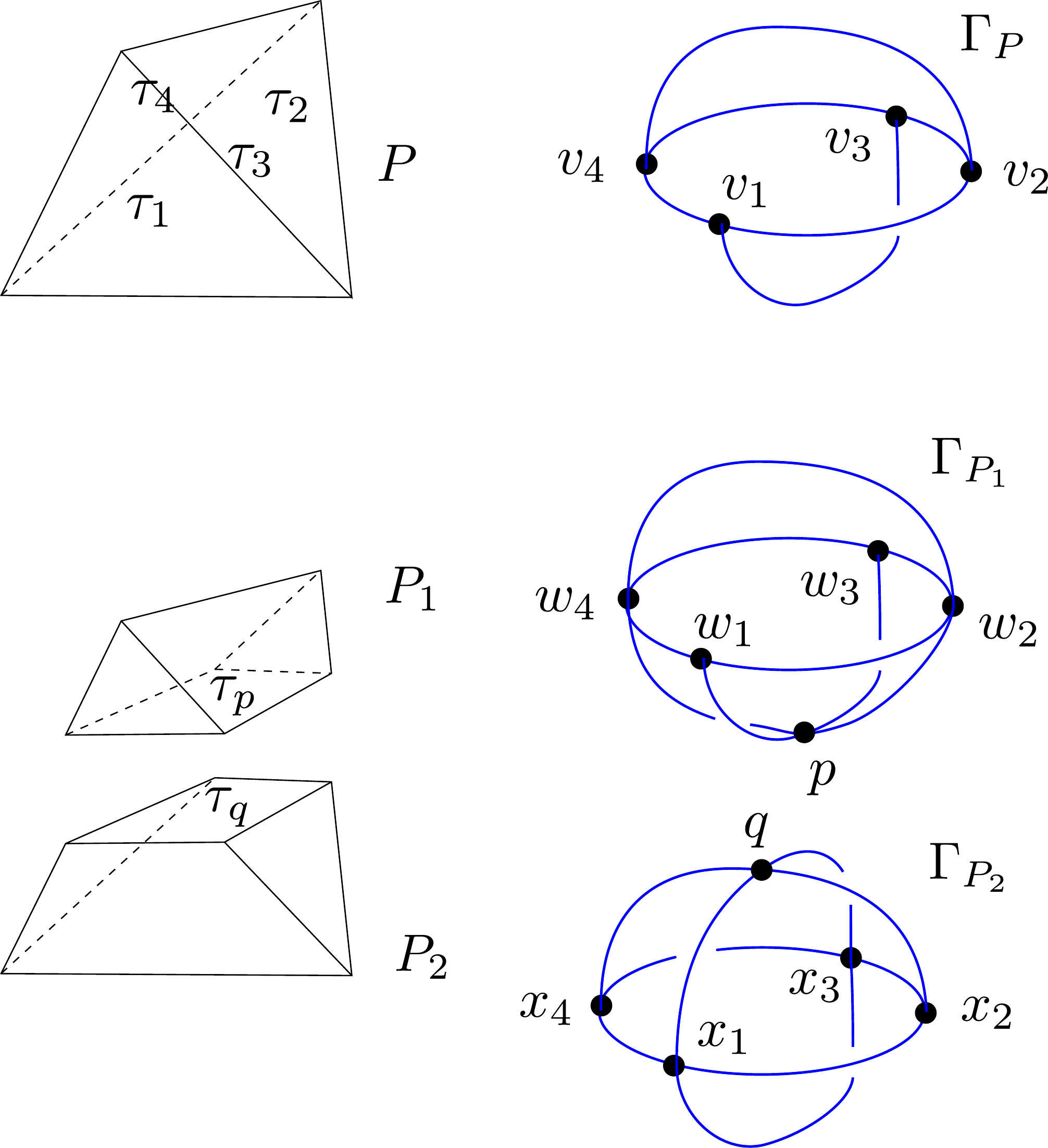}
\caption{Subdivision of a polyhedron $P$ into two $P_1$ and $P_2$, by cutting with a hyperplane. The corresponding boundary graphs get changed correspondingly. This is a $3d$ representation of the $4d$ construction. All four faces of $P$ are subdivided, so in the boundary graph, all four nodes lie on the equator (and get doubled in the cutting process). The new links connecting these four to $v_1$ and $v_2$, respectively, have opposite orientation (not depicted).}\label{Fig:Subdivision_01}
\end{center}
\end{figure}

Now the construction of $\Gamma_{P_1}$ can be achieved as follows: Consider $\Gamma\cap H$, which gives a graph (with open ends) on the northern hemisphere of $S^3$. Add to this another node $p$ at the south pole (corresponding to the 3-face $\tau_p=\partial H\cap P_1$), and connect it to those nodes which lie on the equator $E$. Also, extend every link which passed from the northern to the southern hemisphere (and which now ends on the equator) to the south pole. Then the new links (those from the equator to the south pole) are equipped with an orientation, and the corresponding $B_f$. Denote the nodes of $P_1$, which correspond to $H_1,\ldots H_k$, by $w_1,\ldots,w_k$, while the nodes $v_{k+1},\ldots v_m$ are identical to those of $P$. 

The construction of $\Gamma_{P_2}$ runs along the same lines, where the new node $q$, corresponding to the new 3-face $\tau_q=\partial(-H)\cap P_2$, is placed on the north pole, and all new orientations of faces are chosen to be opposite to the corresponding ones in $P_1$ (see figure \ref{Fig:Subdivision_01}). Denote the nodes of $P_2$ which correspond to $H_1,\ldots,H_k$ by $x_1,\ldots,x_k$, while those corresponding to $H_{ m+1},\ldots, H_n$, which are identical to those of $P$, as $v_{m+1},\ldots,v_n$.

With these conventions, it is clear that $\Gamma_{P_1}\#_{p,q}\Gamma_{P_2}$ can be constructed. However, it is not the same as the original $\Gamma$. The reason is that  in $\Gamma_{P_1}\#_{p,q}\Gamma_{P_2}$, every node which corresponds to $\tau_i$, $1\leq i \leq k$, appears twice, as does every link between two such nodes. The reason is that, after glueing the pieces $P_1$ and $P_2$ back together, the resulting polytope is not quite $P$, but $P$ where every 3-face $\tau_1,\ldots \tau_k$ has been trivially subdivided, so that it counts as two distinct 3-faces.

In particular, the graph $\Gamma_{P_1}\#_{p,q}\Gamma_{P_2}$ does not contain the nodes $v_1,\ldots, v_n$, but rather $w_1,\ldots, w_k, x_1,\ldots, x_k, v_{k+1},\ldots v_n$. Each $w_i$ is connected to $x_i$ for $1\leq i\leq k$. Since the associated 3-faces are the result of cutting a 3-face (dual to $v_i$ in $\Gamma$) into two, the bivectors of both $w_i$ and $x_i$ lie in the same affine  $3d$-subspace which contains $\tau_i$. Hence, the conditions for the move in figure \ref{Fig:MergeNodes} are satisfied, and we can merge the two nodes to a $v_i$. If there is a link $\ell$ between $w_i$ and $w_j$, then there is of course also one $\ell'$ between $x_i$ and $x_j$. Hence, after merging the nodes, we have two links from $v_i$ to $v_j$. These can be merged to one link, and since the bivectors $B_\ell$and $B_{\ell'}$ are the area bivectors of a face $f$ cut in two, we have $B_{\ell}+B_{\ell'}=B_f$. Hence, after merging the nodes and links which were originally on $E$, we are back to the original graph $\Gamma$. \qed

\begin{Corollary}
Let $P$ be a $4$-dimensional convex polytope, embedded in $\mathbb{R}^4$. Let $H$ be a $3d$ half-space which cuts $P$ into two convex $4d$ polytopes $P_1$ and $P_2$. Then
\begin{eqnarray}
I(\Gamma_P)\;=\;I(\Gamma_{P_1})\,+\,I(\Gamma_{P_2}).
\end{eqnarray}
\end{Corollary}
\noindent\textbf{Proof:} This follows immediately from propositions \ref{Prop:EquivalentGraphsHaveEqualInvariants}, \ref{Prop:InvariantBehaviourUnderGlueingOfPolyhedra} and \ref{Prop:GraphBehaviourUnderGlueingOfPolyhedra}.\qed \\

\noindent This means that $I$ associates, to every convex $4d$ polytope, a number in such a way that it is additive under glueing of two convex polytopes to one. The following proposition elucidates the meaning of this number.

\begin{Proposition}\label{Prop:InvariantForSimplices}
Let $P$ be a $4$-simplex. Then $I(\Gamma_P)= V_P$, where $V_P$ is the $4$-volume of $P$.
\end{Proposition}

\begin{figure}[hbt!]
\begin{center}
\includegraphics[scale = 0.75]{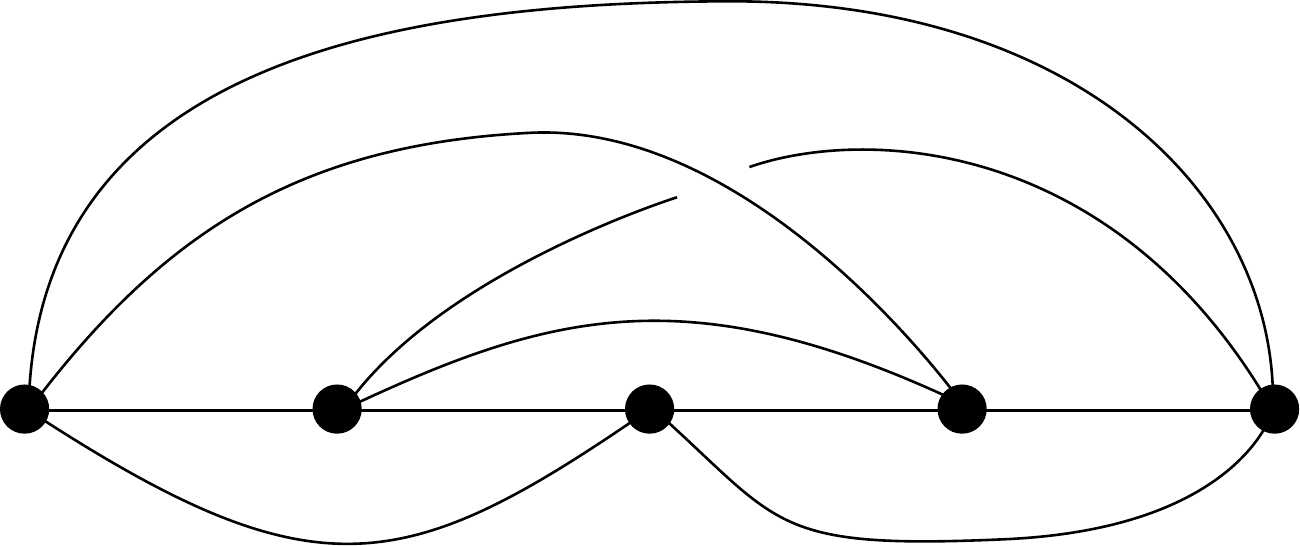}
\caption{The boundary graph of a $4$-simplex.}\label{Fig:4simplex_graph}
\end{center}
\end{figure}

\noindent\textbf{Proof:} This can be shown easily by realizing that there is a projection of $\Gamma_P$ onto the plane with only one crossing (see figure \ref{Fig:4simplex_graph}). 

Each boundary 3-face of $P$ is a tetrahedron, which is the convex hull of all but one vertex of $P$. We label the nodes in $\Gamma_P$ in figure \ref{Fig:4simplex_graph} $N_1,\ldots, N_5$ from left to right, and decree that $N_i$ is the  tetrahedron spanned by all vertices but the $i$-th one. 

The 2-faces of $P$ are triangles, which are spanned by three vertices in $P$. From the figure, one can easily see that the crossing takes place between the triangles $f_1=(235)$ and $f_2=(134)$. Denoting $1$-faces by $(ij)$ if the corresponding $4$-vectors $e_{(ij)}$ going form $i$ to $j$, we can see that (given an orientation of edges such that $\sigma(C)=+1$)
\begin{eqnarray}
B_{f_1}\;=\;\frac{1}{2}e_{(35)}\wedge e_{(32)} ,\qquad B_{f_2}\;=\;\frac{1}{2}e_{(31)}\wedge e_{(34)}.
\end{eqnarray}

\noindent All of these vectors start at vertex $3$, and they span $P$. In particular, from basic geometry, the formula for the $4$-volume of a $4$-simplex is given by
\begin{eqnarray}
V\;=\;\frac{1}{24}*\left( e_{(35)}\wedge e_{(32)}\wedge e_{(31)}\wedge e_{(34)} \right).
\end{eqnarray}

\noindent The claim follows. \qed\\

In other words, the number $I$ associates (up to a factor) the 4-volume to a 4-simplex. So, whenever $I$ associates the volume to two polytopes which can be glued together, $I$ also associates the volume to the result of the glueing. Of course, every $4$-dimensional polytope can be built up from 4-simplices, but there is a subtle caveat we have to clarify, before we can conclude that $I$ indeed associates the 4-volume to every 4-polytope. Namely, in the conditions of proposition \ref{Prop:GraphBehaviourUnderGlueingOfPolyhedra}, all three $P_1$, $P_2$, and $P$ need to be convex. However, by succesively building up a 4-polytope  from 4-simplices, e.g.~by a triangulation, generically intermediate steps will be non-convex. So, we need to show that every 4-polytope can be successively built up from smaller convex pieces, starting with 4-simplices, such that each intermediate step is also convex. This is what we will establish in what follows. 

\begin{Definition}
Assume that, for some convex polytope $P\subset \mathbb{R}^d$, there is a sequence of collections of convex polytopes
\begin{eqnarray}
\{P^{(1)}_1\}\;\to\;
\{P^{(2)}_1,P^{(2)}_2\}\;\to\;
\{P^{(3)}_1,P^{(3)}_2,P^{(3)}_3\}\;\to\;\ldots\;\to\;
\{P^{(m)}_1,\ldots P^{(m)}_m\},
\end{eqnarray}

\noindent such that for one $1\leq m\leq k$ one has 
\begin{eqnarray}
P^{(k)}_l\;=\;P^{(k+1)}_l\,\cup\,P^{(k+1)}_{l+1},
\end{eqnarray}

\noindent while $P_i^{(k)}=P^{(k+1)}_i$ for $i< l$, and $P_i^{(k)} = P_{i+1}^{(k+1)}$ for $i>l$. If all $P^{(m)}_i$, $i=1,\ldots m$ are $d$-simplices, then we call $P$ convex-divisible. 
\end{Definition}

\begin{Proposition}
Every polytope in $d=2$ is convex-divisible.
\end{Proposition}

\begin{figure}[hbt]
\begin{center}
\includegraphics[scale = 0.45]{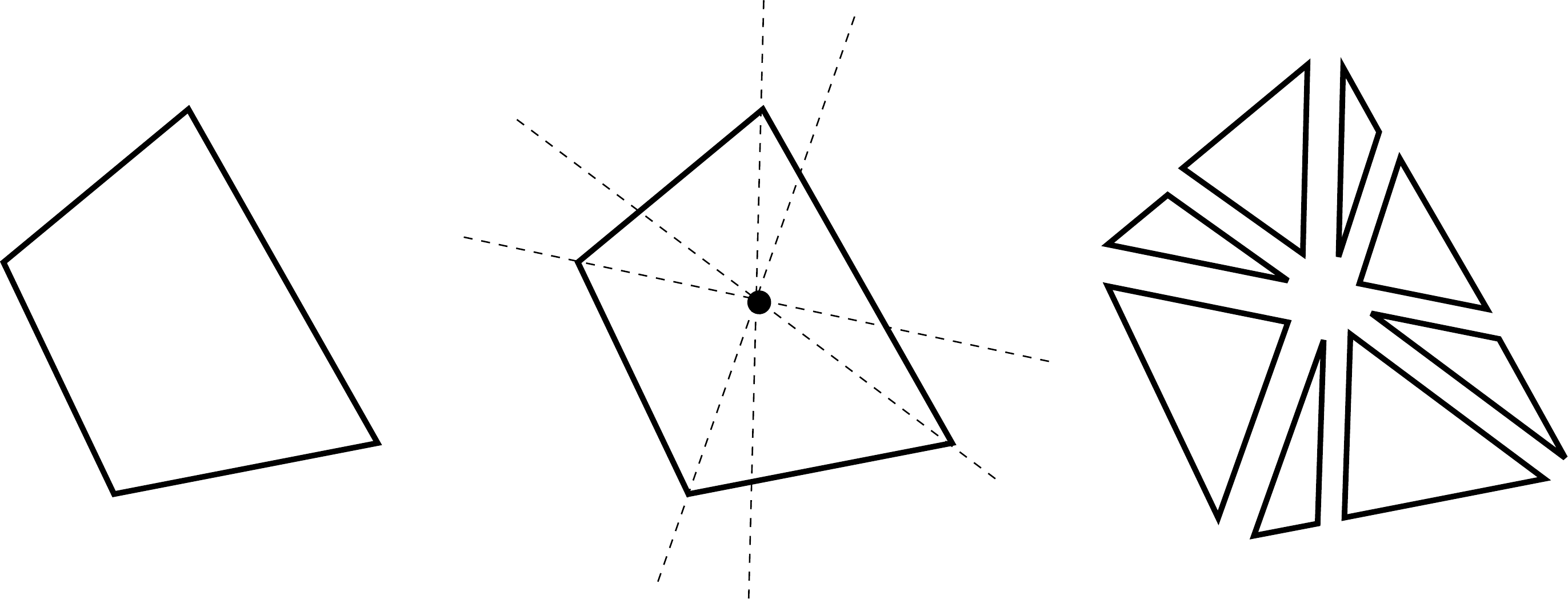}
\caption{Subsequent subdivision of a $4d$ polytope, by the hyperplanes spanned by a vertex $v$ and all $2d$ boundary faces $f$. This is a $2d$ representation of the $4d$ construction, where the boundary faces are 0-dimensional vertices.}\label{Fig:Subdivision_02}
\end{center}
\end{figure}

\noindent\textbf{Proof:}

A polytope in $d=2$ is a polygon with $n$ vertices. Choose a cyclical numbering of these as $v_1$, $v_2,\,\ldots v_n$. If $n=3$, one is done. If $n>3$, then for the first step, cut $P=P^{(1)}_1$ along the line connecting $v_1$ and $v_3$. Then $P^{(2)}_1$ is the triangle with vertices $v_1$,$v_2$, $v_3$, and $P^{(2)}_2$ is the remainder, which has one fewer vertex than $P$. Repeat this process, which stops after $n-3$ steps. \qed

\begin{Proposition}\label{Prop:EveryPolyhedronIsConvexDivisible}
Let $P$ be a convex polytope in $d$ dimensions. If every polytope in dimension $d-1$ is convex-divisible, then $P$ is convex-divisible. 
\end{Proposition}

\noindent\textbf{Proof:} Choose a vertex $v$ in $P$ which is not on the boundary. Every $d-2$-face $f$ of $P$ is the intersection of two boundary $d-1$-polytopes  $\tau_1$ and $\tau_2$ in $P$. The vertices of $f$ lie completely in the $d-1$-dimensional hyperplane spanned by e.g.~$\tau_1$. Since $v$ does not lie in that plane, by construction, $v$ and the vertices of $f$ span a $d-1$-dimensional hyperplane which separates $P$ into two, such that $\tau_1$ and $\tau_2$ lie on either side. 

Let there be $N$ $d-2$-faces in $P$, then the $N$ hyperplanes $H_f$ spanned by $f$ and $v$ successively split the polytope $P$ into $M\leq 2^N$ sub-polytopes $P^{(m)}_i$, $i=1,\ldots, m$. Each of these $P^{(m)}_i$ contains $v$ as a vertex. Every $P^{(m)}_i$ also contains at least one inner point of one of the $d-1$-dimensional boundary faces $\tau$. By construction, $P^{(m)}_i$ then cannot contain any inner point of a different $d-1$ face $\tau'\neq \tau$, since in $P$ these two are separated by at least one $H_f$. 

Hence, each of the $P^{(m)}_i$ is a sub-polytope of the pyramid with $\tau$ as base and $v$ as tip. In fact, it can be generated from that by intersecting this pyramid with half-spaces having $v$ on its boundary. As a result, each $P^{(m)}_i$ is a  pyramid which has $v$ on its tip, and a convex sub-polytope $\hat{\tau}\subset \tau$ as its base.

By our initial condition, $\hat{\tau}$ is convex-divisible, since it is a convex $d-1$-dimensional polytope. The series  of its subdivision determines a subdivision of each $P^{(m)}_i$, by taking the pyramid with tip $v$ (i.e.~the suspension over $v$). The result is a series of $d$-dimensional polytopes being pyramids with bases $d-1$-simplices, and tip $v$. Of course, each of those is a $d$-simplex, so as soon as we have chosen an order of subsequent subdivision of the $P^{(m)}_i$, we are done.

\begin{Corollary}\label{Prop:EveryPolyhedronIsConvexDivisible}
Every convex polytope in any dimension $d\geq 2$ is convex-divisible.
\end{Corollary}

\noindent With this, we finally have everything we need to prove the central claim of this article.

\begin{Lemma}
For any convex 4-polytope $P$, we have that
\begin{eqnarray}
I(\Gamma_P)\;=\;V_P,
\end{eqnarray}
\noindent where $V_P$ is the 4-volume of $P$.
\end{Lemma}
\noindent\textbf{Proof:} By corollary \ref{Prop:EveryPolyhedronIsConvexDivisible}, we can choose a subsequent subdivision of $P$ into convex sub-polytopes, until we arrive a collection of 4-simplices. By using propositions \ref{Prop:InvariantBehaviourUnderGlueingOfPolyhedra} and \ref{Prop:InvariantForSimplices}, the claim follows.\qed

\section{Summary}

In this article we have delivered a proof for the formula (\ref{Eq:CentralFormula}), which relates the volume $V_P$ of a $4d$ polytope $P$ with its $2d$ bivectors $B_f$, and the crossings $C$ in its boundary graph. Besides its geometrical meaning, the quantization of this formula allows to add a cosmological constant term to the Euclidean signature EPRL-FK Spin Foam model, without resorting to quantum deforming the Hilbert spaces, which made the renormalization of the asymptotical formula accessible in a truncated setting \cite{Bahr:2016hwc, Bahr:2017klw, Bahr:2018gwf}. Also, it allows for a formulation of the quadratic volume simplicity constraint, which is not properly imposed in the KKL extension of the EPRL-FK model \cite{Kaminski:2009fm, Belov:2017who,Bahr:2017klw}. Whether this constraint is sufficient to allow for geometric reconstruction of a $4d$ geometry from the boundary state is still open, and it appears that a linear version of the volume simplicity constraint might be suitable to achieve this \cite{Belov:2017who}. 

The proof relied on the convexity of the polytope $P$. However, geometrically it seems feasible to assume that the formula is true even for non-convex polytopes, as long as the boundary is homeomorphic to $S^3$, i.e.~the polytope is simply-connected. In particular, formula (\ref{Eq:GraphBehaviourUnderGlueingOfPolyhedra}), which prescribes the behaviour of the invariant under glueing, is certainly true also for non-convex polyhedra. The main technical difficulty lies in the exact definition of non-convex polytope, of which there are several inequivalent ones, and a mathematical description which does not rely on the intersection of half-spaces, which only works for the convex case. 

This might be of interest, since non-convex polytopes also appear in the asymptotic analysis of the EPRL-FK model, if $3d$ boundary data admits non-convex glueing in $4d$. The question remains whether these should be suppressed in the path integral or not. A version of the volume simplicity constraint (in particular a linear one, which amounts to $4d$-closure of $3d$ normals) might help in this regard. We aim at returning to this point in another publication.

\section*{Acknowledgements}
The author is indebted to John Barrett and Nathan Bowler for helpful discussions. This work was funded by the project BA 4966/1-1 of the German Research Foundation (DFG).

\bibliography{VolumeBib}

\begin{thebibliography}{10}

\bibitem{thomasbook}
T.~Thiemann, {\em Modern {C}anonical {Q}uantum {G}eneral {R}elativity}.
\newblock Cambridge: Cambridge University Press, 2008.

\bibitem{cbook}
C.~Rovelli, {\em Quantum gravity}.
\newblock Cambridge: Cambridge University Press, 2004.

\bibitem{Perez:2012wv}
A.~Perez, ``{The Spin Foam Approach to Quantum Gravity},'' {\em Living Rev.
  Rel.}, vol.~16, p.~3, 2013.

\bibitem{francescabook}
C.~Rovelli and F.~Vidotto, {\em {Covariant Loop Quantum Gravity}}.
\newblock Cambridge Monographs on Mathematical Physics, Cambridge University
  Press, 2014.

\bibitem{Ambjorn:2013apa}
J.~Ambjorn, A.~Gorlich, J.~Jurkiewicz, and R.~Loll, ``{Causal dynamical
  triangulations and the search for a theory of quantum gravity},'' {\em Int.
  J. Mod. Phys.}, vol.~D22, p.~1330019, 2013.

\bibitem{Bombelli:1987aa}
L.~Bombelli, J.~Lee, D.~Meyer, and R.~Sorkin, ``{Space-Time as a Causal Set},''
  {\em Phys. Rev. Lett.}, vol.~59, pp.~521--524, 1987.

\bibitem{Oriti:2007qd}
D.~Oriti, ``{Group field theory as the microscopic description of the quantum
  spacetime fluid: A New perspective on the continuum in quantum gravity},''
  {\em PoS}, vol.~QG-PH, p.~030, 2007.

\bibitem{Plebanski:1977zz}
J.~F. Plebanski, ``{On the separation of Einsteinian substructures},'' {\em J.
  Math. Phys.}, vol.~18, pp.~2511--2520, 1977.

\bibitem{Horowitz:1989ng}
G.~T. Horowitz, ``{Exactly Soluble Diffeomorphism Invariant Theories},'' {\em
  Commun. Math. Phys.}, vol.~125, p.~417, 1989.

\bibitem{Reisenberger:1996pu}
M.~P. Reisenberger and C.~Rovelli, ``{'Sum over surfaces' form of loop quantum
  gravity},'' {\em Phys. Rev.}, vol.~D56, pp.~3490--3508, 1997.

\bibitem{Baez:1999sr}
J.~C. Baez, ``{An Introduction to spin foam models of quantum gravity and BF
  theory},'' {\em Lect. Notes Phys.}, vol.~543, pp.~25--94, 2000.

\bibitem{Barrett:2009gg}
J.~W. Barrett, R.~J. Dowdall, W.~J. Fairbairn, H.~Gomes, and F.~Hellmann,
  ``{Asymptotic analysis of the EPRL four-simplex amplitude},'' {\em J. Math.
  Phys.}, vol.~50, p.~112504, 2009.

\bibitem{Barrett:2009mw}
J.~W. Barrett, R.~J. Dowdall, W.~J. Fairbairn, F.~Hellmann, and R.~Pereira,
  ``{Lorentzian spin foam amplitudes: Graphical calculus and asymptotics},''
  {\em Class. Quant. Grav.}, vol.~27, p.~165009, 2010.

\bibitem{Conrady:2008mk}
F.~Conrady and L.~Freidel, ``{On the semiclassical limit of 4d spin foam
  models},'' {\em Phys. Rev.}, vol.~D78, p.~104023, 2008.

\bibitem{Barrett:1997gw}
J.~W. Barrett and L.~Crane, ``{Relativistic spin networks and quantum
  gravity},'' {\em J. Math. Phys.}, vol.~39, pp.~3296--3302, 1998.

\bibitem{Engle:2007wy}
J.~Engle, E.~Livine, R.~Pereira, and C.~Rovelli, ``{LQG vertex with finite
  Immirzi parameter},'' {\em Nucl. Phys.}, vol.~B799, pp.~136--149, 2008.

\bibitem{Freidel:2007py}
L.~Freidel and K.~Krasnov, ``{A New Spin Foam Model for 4d Gravity},'' {\em
  Class. Quant. Grav.}, vol.~25, p.~125018, 2008.

\bibitem{CubulationSpinFoamThiemann2008}
A.~Baratin, C.~Flori, and T.~Thiemann, ``{The Holst Spin Foam Model via
  Cubulations},'' {\em New J. Phys.}, vol.~14, p.~103054, 2012.

\bibitem{Baratin:2011hp}
A.~Baratin and D.~Oriti, ``{Group field theory and simplicial gravity path
  integrals: A model for Holst-Plebanski gravity},'' {\em Phys. Rev.},
  vol.~D85, p.~044003, 2012.

\bibitem{Kaminski:2009fm}
W.~Kaminski, M.~Kisielowski, and J.~Lewandowski, ``{Spin-Foams for All Loop
  Quantum Gravity},'' {\em Class. Quant. Grav.}, vol.~27, p.~095006, 2010.
\newblock [Erratum: Class. Quant. Grav.29,049502(2012)].

\bibitem{Bahr:2015gxa}
B.~Bahr and S.~Steinhaus, ``{Investigation of the Spinfoam Path integral with
  Quantum Cuboid Intertwiners},'' {\em Phys. Rev.}, vol.~D93, no.~10,
  p.~104029, 2016.

\bibitem{Bahr:2017ajs}
B.~Bahr and V.~Belov, ``{On the volume simplicity constraint in the EPRL spin
  foam model},'' 2017.

\bibitem{Belov:2017who}
V.~Belov, ``{Poincar\'e-Pleba\'nski formulation of GR and dual simplicity
  constraints},'' 2017.

\bibitem{Dittrich:2008va}
B.~Dittrich and S.~Speziale, ``{Area-angle variables for general relativity},''
  {\em New J. Phys.}, vol.~10, p.~083006, 2008.

\bibitem{Freidel:2010aq}
L.~Freidel and S.~Speziale, ``{Twisted geometries: A geometric parametrisation
  of SU(2) phase space},'' {\em Phys. Rev.}, vol.~D82, p.~084040, 2010.

\bibitem{Freidel:2013bfa}
L.~Freidel and J.~Ziprick, ``{Spinning geometry = Twisted geometry},'' {\em
  Class. Quant. Grav.}, vol.~31, no.~4, p.~045007, 2014.

\bibitem{Bahr:2018ewi}
B.~Bahr and G.~Rabuffo, ``{Deformation of the EPRL spin foam model by a
  cosmological constant},'' 2018.

\bibitem{Bahr:2018gwf}
B.~Bahr, G.~Rabuffo, and S.~Steinhaus, ``{Renormalization in symmetry
  restricted spin foam models with curvature},'' 2018.

\bibitem{Han:2011aa}
M.~Han, ``{Cosmological Constant in LQG Vertex Amplitude},'' {\em Phys. Rev.},
  vol.~D84, p.~064010, 2011.

\bibitem{Haggard:2014xoa}
H.~M. Haggard, M.~Han, W.~KamiÅ„ski, and A.~Riello, ``{SL(2,C) Chern-Simons
  Theory, a non-Planar Graph Operator, and 4D Loop Quantum Gravity with a
  Cosmological Constant: Semiclassical Geometry},'' {\em Nucl. Phys.},
  vol.~B900, pp.~1--79, 2015.

\bibitem{Fairbairn:2010cp}
W.~J. Fairbairn and C.~Meusburger, ``{Quantum deformation of two
  four-dimensional spin foam models},'' {\em J. Math. Phys.}, vol.~53,
  p.~022501, 2012.

\bibitem{Han:2010pz}
M.~Han, ``{4-dimensional Spin-foam Model with Quantum Lorentz Group},'' {\em J.
  Math. Phys.}, vol.~52, p.~072501, 2011.

\bibitem{Dittrich:2016typ}
B.~Dittrich and M.~Geiller, ``{Quantum gravity kinematics from extended
  TQFTs},'' 2016.

\bibitem{Reidemeister_Knots}
K.~Reidemeister, ``{Elementare Begruendung der Knotentheorie},'' {\em
  K.~Abh.Math.Semin.Univ.Hambg.}, vol.~4, 1927.

\bibitem{AlexanderBriggs}
J.~W. Alexander and G.~B. Briggs, ``{On types of knotted curves},'' {\em
  Ann.~of Math.}, vol.~2, 1926.

\bibitem{GrossTuckerGraphBook}
J.~L. Gross and T.~W. Tucker, {\em {Topological Graph Theory}}.
\newblock Wiley Interscience, 1987.

\bibitem{MinkowksiTheorem}
H.~Minkowski, ``{Allgemeine Lehrsaetze ueber die convexen Polyeder},'' {\em
  Nachrichten v.~d.~Gesellschaft d.~Wissenschaften zu Goettingen}, 1897.

\bibitem{Bahr:2016hwc}
B.~Bahr and S.~Steinhaus, ``{Numerical evidence for a phase transition in 4d
  spin foam quantum gravity},'' {\em Phys. Rev. Lett.}, vol.~117, no.~14,
  p.~141302, 2016.

\bibitem{Bahr:2017klw}
B.~Bahr and S.~Steinhaus, ``{Hypercuboidal renormalization in spin foam quantum
  gravity},'' {\em Phys. Rev.}, vol.~D95, no.~12, p.~126006, 2017.

\end{thebibliography}
\bibliographystyle{ieeetr}

\end{document}